\documentclass{appolb}
\usepackage{epsfig}
\usepackage{graphics}

\def\lessim{\lower.5ex\hbox{$\; \buildrel < \over \sim \;$}}
\def\gtrsim{\lower.5ex\hbox{$\; \buildrel > \over \sim \;$}}
\newcommand{\nc}{\newcommand}
\nc{\ds}{\displaystyle}        \nc{\ts}{\textstyle}
\nc{\rf}[1]{Fig.\,\ref{#1}}    \nc{\rt}[1]{table\,\ref{#1}}
\nc{\req}[1]{Eq.\,(\ref{#1})}  \nc{\eps}{\varepsilon}
\nc{\beq}{\begin{equation}}     \nc{\beql}[1]{\begin{equation}\label{#1}}
\nc{\eeq}{\end{equation}}        
\nc{\beqa}{\begin{eqnarray}}   \nc{\eeqa}{\end{eqnarray}}       
\nc{\bfi}{\begin{figure}}       \nc{\efi}{\end{figure}}

\begin{document}
\title{Strangeness and Quark--Gluon Plasma%
}
\author{Jan Rafelski
\address{Department of Physics, University of Arizona, Tucson, AZ 85721, USA}
}
\maketitle
\begin{abstract}
I review the foundational motivation which led us to the ultra relativistic  heavy ion collision research at SPS, RHIC and now LHC: the quantum vacuum structure;  the deconfined nature of quark-gluon plasma (QGP)  phase filling the Universe for the first 30$\mu$s after the big-bang;  the origin of stable matter mass; and of the origin of flavor. The special role of strangeness enhancement and strange antibaryon signature  is highlighted. It is shown how  hadron production can be used to determine the properties of  QGP, and how the threshold energy for QGP formation is determined. \end{abstract}

\section{Pillars of the QGP/RHI collisions research program}
\subsection{Create the primordial Quark Universe in Laboratory} 
Relativistic heavy ion (RHI) collisions recreate  the high energy density conditions prevailing in the Universe before pre-matter: quarks, gluons,  froze into individual hadrons at about 30$\,\mu$s  after the big-bang. This is the era we can call big-bang matter creation (BBMC). We know now that  the QGP-Universe hadronization  led to a nearly matter-antimatter symmetric state.  The Universe today is filled primarily with the `ash' of the ensuing matter-antimatter annihilation, e.g.  background photons and neutrinos, and a tiny $10^{-9}$ residual matter asymmetry fraction. There is a small but finite chance that by conducting RHI experiments, we will unravel the mechanism of BBMC matter--antimatter asymmetry and understand its origin, and thus, understand the deep riddle of matter stability. 

The understanding of the quark Universe deepens profoundly the reach of our understanding of the properties and evolution of the Universe. The   precision  microwave background studies explore the conditions in the Universe at temperatures near  the scale of $T=0.25$ eV where hydrogen recombines and photons can  move freely and  the era of observational cosmology begins. Another factor 30,000 into the primordial depth of the Universe expansion, we reach the big-bang nuclear synthesis stage  occurring at the scale of  $T\simeq 10$ keV. A further  factor 30,000   increase of  temperature   is needed to reach the BBMC stage at which the hadronization of quark Universe occurs at Hagedorn temperature $T\simeq 160$ MeV. 

\subsection{Explore the nature of the quantum vacuum: Einstein's \ae ther}
The vacuum state determines  prevailing fundamental laws of nature. Within the standard model, the   nature of  particles and their  interactions is determined by the transport properties of the vacuum state. The existence of a structured quantum vacuum as the carrier of the laws of physics was anticipated by Lorentz  and Einstein, they called it \ae ther. Writing to Lorentz in November, 1919 Einstein says: {\em It would have been more correct if I had limited myself, in my earlier publications, to  emphasizing only the non-existence of an \ae ther velocity, instead of arguing the total non-existence of the \ae ther,\ldots}.   Within a year Einstein writes~\cite{sidelights} {\em \ldots space is endowed with physical qualities; in this  sense,  therefore, there exists an \ae ther\ldots.   According to the general theory of relativity space without \ae ther is unthinkable; for in such space  there not only would be no propagation of light, but also no  possibility of existence for standards of space and time (measuring-rods  and  clocks), nor therefore any space-time intervals in the physical  sense. But this \ae ther may not be thought of as endowed with the quality characteristic of ponderable media, as (NOT) consisting of  parts   which may be tracked through time. The idea of motion may not  be applied to it.}  

In the quark--gluon plasma state of matter, we fuse and dissolve nucleons in the primordial \ae ther state, different in its structure and properties from our experience. We can now explore questions such as: what is the velocity of light in the new QGP vacuum state? What makes color charge mobile?

\subsection{The origin of mass of matter: deconfinement} 
The confining quark vacuum state is contributing 99.9\% of mass of matter surrounding us. The  Higgs mechanism applies to the remaining 0.1\%. Only the very heavy, and unstable, quarks are strongly connected to the Higgs sector.  The   quantum zero-point energy  of localized light quarks is believed to govern the mass of matter; we demonstrate this by setting quarks free  in  laboratory experiments involving collisions of large nuclei at relativistic energies. In the collision, several  reaction steps occur:\\
\indent 1) formation of  the primary fireball; a momentum equipartitioned partonic  phase comprising in a limited space-time domain the final state entropy;\\ 
\indent 2) the cooking of the energy content of the hot matter fireball  towards the particle  yield (chemical) equilibrium in a hot perturbative quark--gluon  plasma phase --- this is the quark--gluon plasma liquid (QGP) --- a drop of the matter that filled the  universe up to about 30$\mu$s;\\
\indent 3) emergence near to the phase boundary of  {\em transient}  massive effective quarks and  disappearance of free gluons; this phase cannot be in chemical equilibrium if entropy, energy,  baryon number and strangeness are to be conserved;\\
\indent 4) hadronization, that is combination of  effective and strongly interacting $u,d,s, \bar u,\bar d$ and $ \bar s$ quarks  and   anti-quarks   into the final state  hadrons, with  the yield probability weighted by accessible phase space. 

In this manner, experiments probing quark deconfinement are leading to abundant hadron production. Much of time and effort is dedicated to cataloging  experimental outcome: analyzing the multiparticle debris in search of new physics, as I will report in more detail below.

\subsection{What is flavor} 
The matter we see and touch is made solely of first flavor family ($u,\ d,\ e,$ $\nu_e$).  In RHI collision experiments we form the new state of matter comprising  hundreds if not thousands (at the LHC) elementary particles of the unstable 2nd family ($s,\ c,\ \mu,\ \nu_\mu$): this is perhaps the  only  laboratory environment where as much as one-third of energy has been converted into particles of  the 2nd flavor family.  For this reason, I believe that this is perhaps one and the only opportunity we have to study and hopefully unravel the secret of flavor.

\section{QGP probes}
\subsection{Remote sensing the QGP}
At RHIC and LHC stored beams of relativistic nuclei,  each  about 6 fm in size, pass through each other many times before they accidentally hit each other, leading in some small fraction of events to near head-on collisions. The time available to travel across the nuclear volume which is the reaction zone is in range of $10^{-22}$--$10^{-23}$s. The  extraordinary experimental challenge is to learn how to detect and understand physics occurring in this incredibly short blink of time. I will focus, here, on the one probe that has proved itself: strangeness. 

To fully appreciate how special the role of strangeness is, it is helpful to first consider the key pros and cons of using more conventional probes.  Imagine a method similar to observation of the Sun or the early Universe, i.e., photons that emerge from the reaction as a probe the conditions prevailing. However, photons are weakly coupled to matter, on the short time scale explored only a few will be produced, and the background photons   from decay $\pi^0\to \gamma \gamma$ are vastly dominant (note that $\gamma$ here is not to be confounded with later use of $\gamma$ as a fugacity). A virtual photon with $q^2\ne 0$, a dilepton, has a better chance of success, as the natural backgrounds are expected to be smaller. However, the difficulty of a relatively weak electromagnetic coupling to dense matter remains, while at the same time it turned out that it is difficult to understand the dilepton background sourced in hadron phase. 

On the other end of interaction strength are probes that are strongly coupled. For example, high energy individual quarks `partons' can interact with matter and as result, with distance traveled the energy is dissipated, as one says `thermalized'. Since at the production point a second high energy quark was produced, we can deduce from `jet' asymmetry that the dense matter we form in RHI collisions is very opaque, and with some effort we can quantify the strength of such an interaction. Like  $J/\Psi(c\bar c)$ suppression this method offers a quick peak into the QGP soup but it lacks  specificity inherent in a probe that has many facets. Moreover, even the inference using this observable that we formed a new phase of matter, QGP,  is entirely theoretical and comprises one single piece of output information, the strength of interaction of a parton, or $J/\Psi$, in the dense matter. 

\subsection{Strangeness as signature of QGP}
The events accompanying the discovery and development of strangeness signature of QGP more than 30 years ago have been reported~\cite{Rafelski:2007ti}. This is a short recapitulation  of three important issues raised which since have seen a long and tedious  development:
\begin{enumerate}
\item  The  chemical equilibration  of strange quarks in QGP --- with the yield increasing with more extreme initial conditions and larger size of the QGP   probes the  earliest stages of the QGP formation;
\item The combinant quark hadronization which offers an image of the late stages of dense QGP matter;
\item The comparison with scaled nucleon-nucleon reactions and with  other non QGP mechanisms allows a theory independent assessment of multi-strange  hadrons and, in particular, antibaryons as a signature of QGP. 
\end{enumerate}
Strangeness has proved, over the past two decades, to be a functional remote sensing probe that differentiates stages of collision and properties of matter formed. The reason is that strange quarks are  strongly interacting particles  from the 2nd flavor generation, and thus,  remain relatively weakly coupled to matter made of the 1st generation. The medium strength of the interaction and small background offers a workable compromise. 

\subsection{Strangeness observables}
\underline{Strangeness $s$ to entropy $S$ ratio $s/S$} enhancement  is the `deepest' probe: both $S$ and $s$ are produced early in the reaction, $S$ when colliding parton matter thermalizes (how that happens is not understood) and $s$ in parallel as now thermal gluons seek to equilibrate with thermal quarks. In a chemically equilibrated QGP, this ratio is entirely controlled by the ratio of strange degrees of freedom to all, with important corrections due to the ratio $m_s/T$ of strange quark mass to prevailing temperature. One can argue that this ratio is preserved while QGP breaks apart: production, and equally, destruction of strange quark pairs is slow at this stage, and disintegration of an entropy dense phase into an entropy thin phase usually does not lead to production of additional entropy. Thus, when we measure $s/S$ in the final state, we look at collision stage at the time of initial thermalization of parton energy. 

\underline{Strange antibaryons} are produced when the unusually abundant $\bar s=s$ antiquarks seek to bind to other quarks in hadronization process. Because $\bar s q$ Kaons are relatively heavy, it is cheaper to emerge with a large abundance of multi-strange baryons and antibaryons, and many hidden strangeness mesons ($\eta, \eta',\phi$). Usually, multi-strange antibaryons and $\phi$  are difficult to make as to put them together, we need to create all the valance quarks separately, while in QGP-soup they are readily available. For this reason, strange antibaryons and $\phi$ are  probes of QGP presence at hadronization.

\underline{Strange resonances}, Strange hadrons are  stable  on the scale of the collision, with a lifespan  in comparison to the reaction time  larger than the age of Universe expressed in years. However, strange resonances, the excited states of strange hadrons,  are also produced abundantly in the hadronization process. Strange resonances live on time scales comparable to all other strongly interacting processes. Therefore they help quantify the duration of hadronization and also characterize the after-life of matter.

\underline{Heavy flavor ($c,b$)} is expected to bind preferably with strangeness creating novel forms of matter even more sensitive to the source properties --- we are all looking forward  to forthcoming results as this new physics  begins to be explored.

\section{Strangeness and Quark--Gluon Plasma}
\subsection{Strangeness production in QGP}
The QGP state emerging from initial parton collisions can reach kinetic and even flavor chemical equilibrium. We can measure this observing the produced particle yields; the final state hadrons are the vapor of a boiling quark--gluon drop. This hadron `vapor'  carries complete information about the boiling state of the hot quark drop. The high strangeness abundance in the QGP drop, along with the resultant recombinant high strange antibaryon yields, offers an opportunity  to demonstrate  formation of the  deconfined quark--gluon matter, and provides information about the source bulk  properties.

The production of strangeness is a dynamical time dependent process\footnote{It is inappropriate, in 2011 to speak of particle production when presenting a non-dynamic equilibrium description of their yield,  a presumed `chemical equilibrium abundance' in the phase of matter one considers: QGP or hadron gas.}. In the QGP, strangeness pair production is mainly due to gluon fusion processes, $gg\to s\bar s$~\cite{Rafelski:1982pu}, but light quarks also contribute $q\bar q\to s\bar s$~\cite{Biro:1981zi}. In the local rest frame, the change in density of either strange or anti-strange quarks can be written in terms of momentum distribution averaged reaction cross sections and result in the master equation having the form:
\beq\label{master}
{d \rho_{\bar s,s}\over d \tau} 
=
\frac12 \rho_g^2(t)\,\langle\sigma v\rangle_T^{gg\to s\bar s}
+
\rho_{q}(t)\rho_{\bar q}(t)
\langle\sigma v \rangle_T^{q\bar q\to s\bar s}
-
\rho_{s}(t)\,\rho_{\bar{\rm s}}(t)\,
\langle\sigma v \rangle_T^{s\bar s\to gg,q\bar q}.
\eeq

When the last loss term balances the gain terms, the (chemical) equilibrium yield of strangeness is achieved. In general, the process of momentum equilibration (kinetic equilibrium) is considerably faster as any reaction occurring in the dense QGP phase contributes. The slower particle abundance equilibration is described  by introducing a phase space occupancy parameter, for strangeness $\gamma_s(t)$, but more generally, chemical non-equilibrium abundance can be considered for any component of QGP. To be specific, the quantum phase space distribution which maximizes the entropy at fixed particle yield is:
\beq\label{fermi}
{{d^6N}\over {d^3pd^3x}}\equiv f( p)={g\over (2\pi)^3}\frac 1 {\gamma_i^{-1}\lambda_{i,k}^{-1} e^{E_i/T}\pm 1},
\quad i=q,s,c,b; \quad k=B,S,
\eeq
where $E_i=\sqrt{p^2+m_i^2}$. Note that independent of the values of particle masses $m_i$,  chemical potentials $\mu_{B,S}=T\ln \lambda_{B,S}$ and  phase space occupancy $\gamma_{s,c,b}(t)$, we have for fermions $0<f(\vec p)<{g/(2\pi)^3}$ as required, where $g$ is the statistical degeneracy.

\subsection{Strangeness chemical equilibration in QGP}
\begin{figure}[tb]
\centerline{
\epsfig{width=6cm,figure=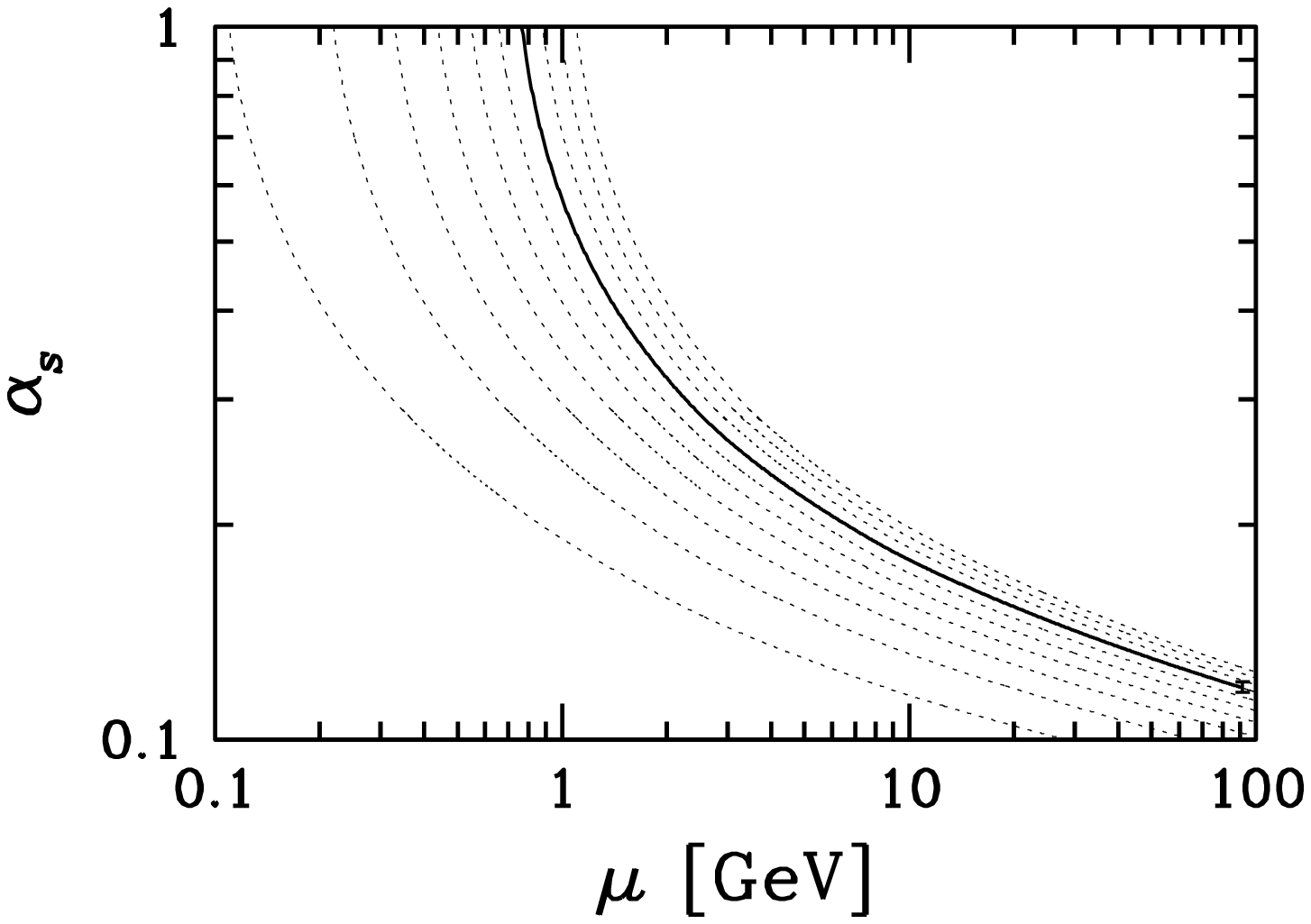}
\epsfig{width=6cm,figure=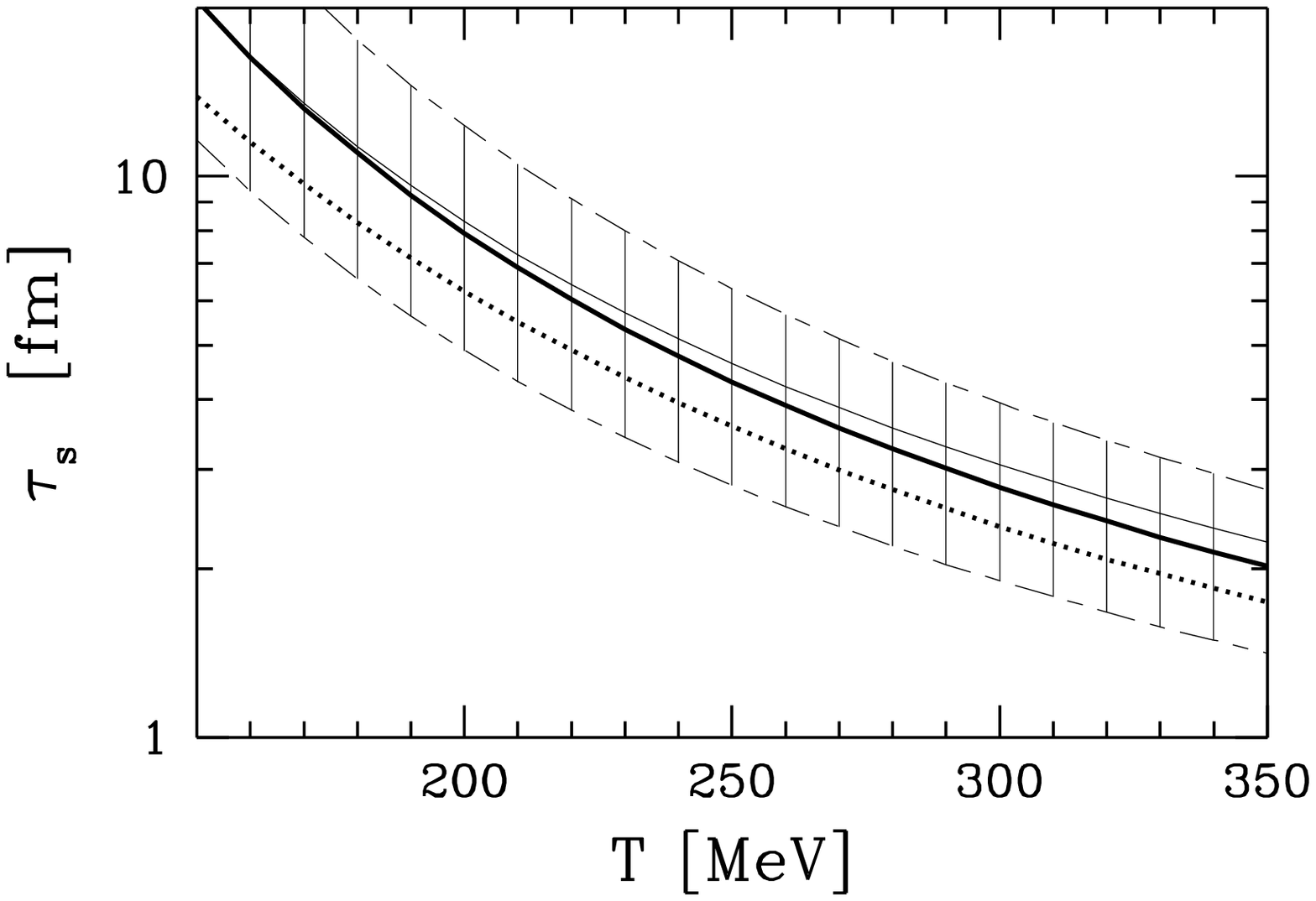}
}
\vskip -.2cm
\caption{\label{alphamu}Left: $\alpha^{(4)}_{\rm s}(\mu)$ as function of energy scale $\mu$ for a variety of initial conditions at $\mu=M_{Z}$. Solid line: $\alpha_{\rm s}(\mu=M_{Z})=0.1182$ (marked by experimental point). MSS scheme through 4th order (superscript (4)) was used in running the coupling strength. Right: relaxation time $\tau_s$ near to chemical equilibrium as function of $T$.  See text for further details.
}
\end{figure}

The pertinent cross sections $\sigma$ are evaluated within the context of perturbative QCD, and much improvement comes when one uses the scale-running QCD parameters, the coupling $\alpha_s(\mu)$ and strange quark mass $m_s(\mu)$. The scale dependent variation of $\alpha_s(\mu)$ is summarized in the figure \ref{alphamu} (left panel). The fat line is constrained to the experimentally measured value  $\alpha_{\rm s}(M_{Z})=0.1182$, other lines show how a tiny change of $\alpha_{\rm s}(M_{Z})$ impacts the strength of coupling towards smaller energy scales.

On the right in figure \ref{alphamu}, the characteristic time is shown that passes while the strangeness production reactions chase the equilibrium strangeness abundance. This `chemical relaxation' time is obtained  by dividing the density that is `chased' $\rho_s^{\rm equilibrium}$ by the rate at which it is chased, which is the production term on the right of \req{master}. We wee the value evaluated  near to equilibrium condition  which is an upper limit: the further one is from equilibrium, the faster the approach to equilibrium occurs. When the initial kinetic equilibrium temperature is near to $T=400$ MeV, strangeness formation is thus very rapid, on the scale of $\tau_s<1$\,fm/c.

The main uncertainty seen on right in   figure \ref{alphamu} is due to  uncertain mass of the strange quark.  Central lines offer two different schemes for running $\alpha_s$, and thus, characterize uncertainty in the understanding of QCD coupling strength. Dotted line is what was used in 1982 with fixed $\alpha_s=0.6$~\cite{Rafelski:1982pu}: $\alpha_s=0.60\pm0.10\pm0.07$ arises at $\mu=0.86\rm{GeV}$.  Not shown are  systematic, uncertainties due to higher order processes not discussed here. However, these are smaller than the mass scale uncertainty.

\subsection{Other quark flavor in QGP}
Strangeness approaches chemical equilibrium yield from below, that is, in general, at any early time in the formation of QGP $\gamma_s(t_0)<1$ and the following evolution is due to thermal reactions described by \req{master}. Different initial conditions apply  for the very heavy charm and beauty quarks. The chemical equilibrium yields  characterized by the factor $(m_{c,b} T)^{3/2}e^{-m_{c,b}/T}$ are very small, and yet smaller when we correlate particles and antiparticles in a small thermal volume. Considering collisions  at rather high energy as is the case at RHIC or LHC one finds that  primary parton collisions provide in general a much higher charm and/or bottom yield than needed to saturate the small chemical equilibrium yield. Thus $c,\ b$  begin their QGP  evolution history with $\gamma_{c,b}(t_0)>1$. Remarkably, the charm and bottom abundance in QGP is not increasing, but decreasing by thermal processes towards the chemical equilibrium.

Heavy flavor production cross sections in lowest order scale with $\sigma\propto \alpha_s^2/m^2$.  Considering smaller running coupling and much larger mass a great reduction in speed of thermal reaction is predicted. For bottom thermal yield equilibration is   negligible, for charm it is at the level of a few percent.  Conversely,  light quarks equilibrate rather rapidly with the even more strongly self coupled gluons and in general can be assumed to follow and define QGP matter properties.

Heavy quark yields  are related to the pre-thermal parton dynamics. However, heavy quarks may acquire through elastic collisions a momentum distribution characteristic of the medium, providing an image of the collective dynamics of the dense quark matter flow.

This discussion shows  that only strangeness yield has the very fine feature of being sensitive to the conditions prevailing in the QGP phase, probing in the yield the most extreme thermal stage. Because strangeness abundance is indisputably driven by gluons, achievement of chemical equilibrium in QGP requires the presence of mobile, free gluons, and thus of deconfinement. Moreover, nearly  20\% of all energy content of QGP is transfered to strangeness when the chemical equilibrium is reached, and empowering the possibility to investigate the riddles of flavor.

\section{Hadronization of QGP}
\subsection{Production of hadrons}
Enrico Fermi was the first to address the multi particle production phenomenon in hadron collisions. He proposed  the hypothesis~\cite{Fer50} that `strong' interactions  saturate the  quantum  production matrix elements. Therefore, pursuant to Fermi's golden rule, the yield of particles is described alone by the relative magnitude of the accessible phase space, and is overall constrained by energy conservation: the original Fermi model was `micro-canonical', that is the accessible  phase space was considered in terms of the available collision energy. The thermodynamic picture was developed in the following 30 years, and has led on to the recognition of the phase transition from hadrons to the deconfined quark--gluon matter~\cite{Hag84}. The Hagedorn temperature $T_H=160$ MeV became the boiling point of QGP.

However, the use of temperature-like parameter $T $, which describes the magnitude of the phase space for the final state hadron particle yield,  does not mean by necessity  that there is an equilibrated hadron gas in the final state comprising a multitude of interacting hadrons. Rather, if a drop of hot QGP  decays into streaming not interacting hadrons, $T$ is a parameter allowing to count particles and assuring within a   good approximation (along with other statistical parameters) on average the conservation of the QGP-drop energy content. Similarly, baryon number content is conserved by use of baryochemical potential, etc.  These parameters, measured by observing hadron yield describe the properties of the particle source.

Thus the hypothesis of an chemical abundance equilibrium condition among hadrons does not apply, hadrons yields follow from hadronization dynamics~\cite{Koch:1986ud}.  When a chemical hadronic equilibrium state is observed, we can presume that QGP was either not formed of for reasons requiring in depth study, the produced hadrons despite relatively low final state density had an opportunity to reequilibrate.

\subsection{Statistical Hadronization}
Our task is to describe precisely a multitude of hadrons by a relatively small set of parameters. This then allows us to characterize the drop of QGP at the time of hadronization. In our view, the key objective is to characterize the source of hadrons rather than to argue about the meaning of parameter values in a religious fashion. For this  procedure to succeed, it is necessary to allow for greatest possible flexibility in characterization of the particle phase space, consistent with conservation  laws and related physical constraints at the time of QGP hadronization. For example,  the  number yield of strange and light quark pairs has to be   nearly preserved during  QGP hadronization. Such an analysis of experimental hadron yield results  requires a significant book-keeping and fitting  effort, in  order to allow for resonances, particle widths, full decay trees and isospin multiplet sub-states. We use  SHARE (Statistical HAdronization with REsonances), a  data analysis program available for public use~\cite{Torrieri:2004zz}, developed as a join project between Tucson and Krakow groups. 

The important parameters of the SHM, which control the relative yields of particles, are the particle specific fugacity factors ${\lambda}$ and the space occupancy factors ${\tilde \gamma}$ (note that tilde differentiates from QGP objects  discussed above). The fugacity is related to chemical potential ${\mu} = T{\ln{\lambda}}$. The occupancy   ${\tilde\gamma}$ is, nearly, the ratio of produced   particles to the number of particles expected in chemical equilibrium. The meson yield with one quark and antiquark is (nearly) proportional to  $\tilde \gamma_q^2$ and the baryon yield to $\tilde \gamma_q^3$ --- to distinguish the flavor of the valance quark content for $u,\,d,\,s,\ldots$,  we need to introduce the factor $\tilde \gamma_s/\tilde \gamma_q$ for each strange or antistrange quark present in a hadron. 

The occupancy parameters for hadrons (marked by `tilde') will not have the same value as the corresponding phase space occupancies in the QGP. The best way to understand this is to assume that we have a completely equilibrated QGP with all quantum charges zero (baryon number, etc) and thus, in QGP all $\lambda_i=1, \gamma_i=1$ . This state   decays into final state hadrons preserving energy, and increasing or preserving entropy, and, number of pairs of strange quarks. Just two parameters such as   $T$ and volume $V$ would not suffice to satisfy these constraints,  on hadron side, we must  introduce $\tilde \gamma_s>1$. The value is above zero because in the  relevant  domain the QGP state in chemical equilibrium contains greater number of strange quark pairs compared to the hadron phase space.

Clearly, when and if we allow $\tilde \gamma_s$ to account for excess of strangeness content, we must also introduce  $\tilde \gamma_q$ to account for a similar excess of QGP light quark content. Only when and if  we fit $T,V,\tilde \gamma_s, \tilde\gamma_q$ to the particle yields, produced by a QGP source, can we  infer the bulk properties to the (QGP) source that produced these particles. There are two more comments due: 1) we do not know all hadronic particles, the incomplete hadron spectrum used in SHM  is reliably absorbed into $\tilde \gamma_s, \tilde\gamma_q$; 2) we do not fit spectra but yields since the dynamics of outflow of matter after collisional compression is hard to control, but integrating spectra (i.e., yields) are not affected. 

Finally, let us remember that the research groups that lack the skill and/or the  will to use $\tilde \gamma_s$, and/or  $ \tilde\gamma_q$ have long recognized the need to introduce $\tilde \gamma_c, \tilde\gamma_b$. How can this be justified? One cannot argue `we know charm and bottom are out of chemical equilibrium' without allowing for   chemical nonequilibrium for all quark flavors. It is the analysis result which decides which flavors are in equilibrium.

\subsection{Hadron source bulk properties}
Among important features built into SHARE is the capability to fully describe the properties of the QGP drop that produces particles analyzed. This is not done  in terms of evaluation of equations of state for given $T, \mu_B,\ldots$  (which would be wrong), but in terms of  produced particles:   each  carries away energy and quantum numbers.  We evaluate and sum  all fractional contributions to the bulk properties from observed and unobserved particles. 

Furthermore, we can use  any of the QGP  bulk properties to constrain fits to particle yield. This is done by fitting aside of particles also the physical bulk properties. This is particularly important when  there are several fit minima, which is not uncommon in a space of 7 parameters.  Then, it helps finding the physical state  if the information can be input that, e.g., the bulk energy density should be roughly 0.5 GeV/fm$^3$. In fact, the myth that SHM is unstable when $\tilde\gamma_q$ is included originates with groups that neglect to classify the solutions for the parameters according to the physical properties of the source. It is regrettable that, at the time of writing of this report, only the SHARE program has these rather easy to achieve capabilities, which are necessary in order to understand the physics of hadron productions and QGP formation within the full nonequilibrium approach.

\begin{figure}[tb]
\centerline{
\epsfig{width=6.9cm,figure=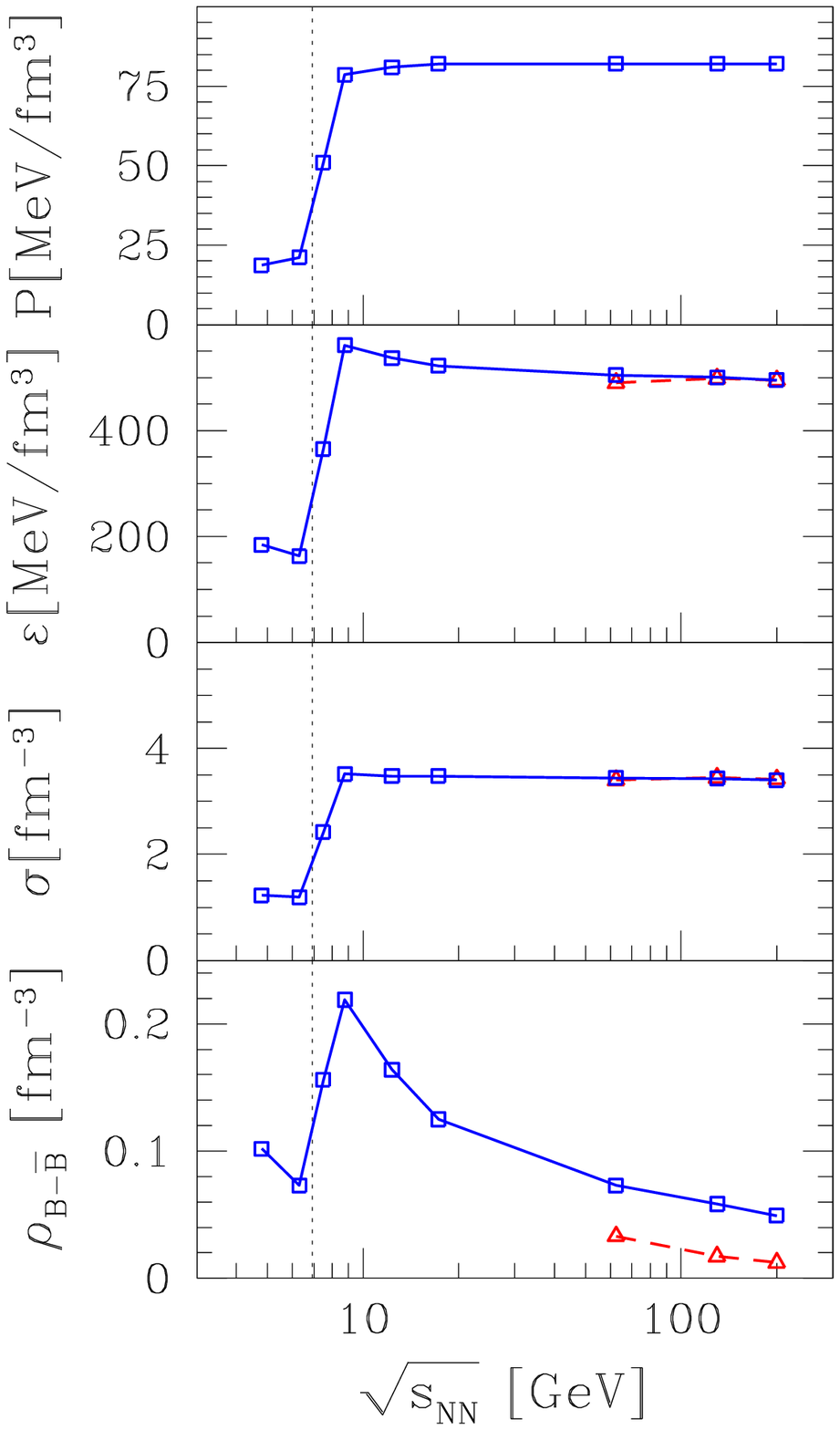}\hspace*{-0.5cm}
\epsfig{width=6.95cm,figure=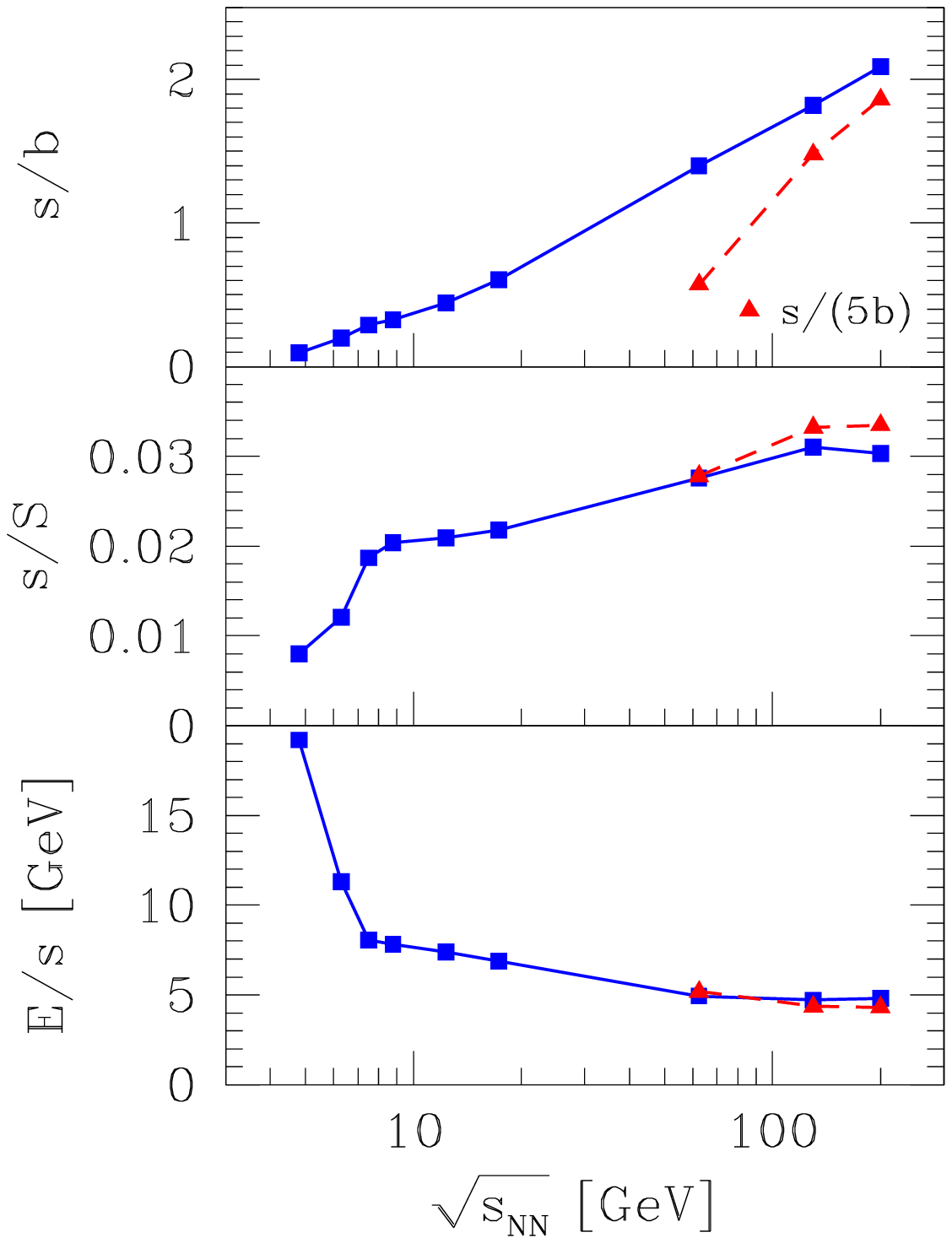}
}
\vskip -.2cm
\caption{\label{Edep} 
The physical properties of hadron source obtained summing contributions of different particles with parameters  fitted to SPS and RHIC $4\pi$ data, presented as function of collision energy. For RHIC range also results related to $dN/dy$ are shown. On left: bulk properties from top to bottom: pressure, energy density, entropy density and baryon density. On right, various ways of seeing strangeness yield increase with collision energy are presented: strangeness per baryon, strangeness per entropy, and, bottom, thermal energy needed to produce a strange quark pair.The analysis is done for  total particle yields. $dN/dy$ results are shown as dashed lines in the RHIC energy range.}
\end{figure}

Let me close this short report with update of the answer to the question posed   by Marek Ga\'zdzicki --- `what is the meaning of `my'   horn in $K^+/\pi^+$ abundance' --- see figure 3 in Ref.~\cite{Grebieszkow:2011gu} in this volume. Is this a signal of the onset of new physics?  Our analysis of these results, which includes at lowest energy the AGS top energy results, and at high energy a wide range of RHIC results, is shown in figure \ref{Edep}, which is an update of our earlier work~\cite{Letessier:2005qe}.

We see that where the peak of the  $K^+/\pi^+$ ratio horn occurs, we find a peak in net baryon density, bottom frame of the figure \ref{Edep} on left: an effect of baryon stopping, at low energy compression of baryon number occurs, and beyond a certain collision energy the finite nuclear size allows a `shoot through'. This  onset of transparency maybe indeed due to the onset of deconfinement at this energy. The argument that a decrease in baryon density is due to expansion is not right, seen that  the energy density $\epsilon$, and the entropy density $\sigma$,  remain constant above a threshold in collision energy. 

In figure \ref{Edep} we show on right a smooth increase of strangeness production: strangeness per baryon is growing smoothly (top frame), and we see  a `knee'  in $s/S$ (middle frame) where baryon density peaks, the change in slopes suggests that entropy production mechanism changed supporting the notion of onset of formation of a new phase of matter. The relative decline in $K^+$ yield  originates in the abundant formation of particles with hidden strangeness such as $\eta,\ \eta',\ \phi$ in the QGP hadronization process: in QGP breakup it is `cheaper' to hide strange quarks in a relatively low mass $\eta$ than to make   two kaons. 

Our analysis thus shows: a) There is an onset of baryon transparency and entropy production at a very narrowly defined collision energy range. b) Beyond this threshold in  collision energy  the hadronization proceeds more effectively into hidden strangeness and strange antibaryons, this depletes $K^+$ yield. c) The universality of hadronization source properties, such as energy density, or entropy density above the same energy threshold, suggest as explanation that a new phase of matter hadronizes. d) The decline in $K^+/\pi^+$ ratio parallels the decline in net baryon density but is unrelated. 

There is little doubt considering these cornerstone analysis results  that beyond Ga\'zdzicki-horn energy threshold we produced a rapidly evaporating (hadronizing) drop of QGP. As the energy is reduced below the threshold,  the state of matter is less clear. We have speculated~\cite{Letessier:2005qe}  that we form deconfined state of massive constituent quarks. This is consistent with  a rapidly rising `open' strangeness yield, and a small increase in entropy, and the  rapidly rising baryon density. Notable is the  more precipitous drop in thermal energy cost to make a strange quark pair, while the bulk properties where from hadrons originate become much less extreme.

\vfill
\section{Conclusions}
Strangeness has proved to be a most useful QGP observable in the entire range of energies explored.  The proposed strangeness enhancement has been observed~\cite{Gazdzicki:1995cr}. The proposed strange antibaryon enhancement has been observed~\cite{Alber:1996mq,Antinori:2006ij}. 

The reports at this meeting on enhancement at LHC are very encouraging --- it is perhaps also a good time to mention that all LHC hadron production results including multi-strange baryons and antibaryons are very well fitted with SHARE and result in parameters and bulk properties that align well with the results seen at SPS and RHIC with the possible exception that the  normalization volume is 30--50\%  larger than is predicted by HBT systematics. Similarly, the ratio $s/S$ is about 5\% smaller than at RHIC. All this is very interesting, however,  analysis of LHC-ion hadron yields  deserves its own publication which is forthcoming. 

Similarly our hadronization parameters predict correctly the low $K^*$ yield reported by NA49, there is no need or opportunity to invoke novel mechanisms of $K^*$ absorption. Chemical non-equilibrium model works at SPS, RHIC, LHC, and with it strangeness signature of QGP comes of age.

Strangeness experimental results fulfill   our expectations: they offer a resounding confirmation of  fast hadronization of quark--gluon plasma in that we observe $m_\perp$ spectra that are the same up to normalization for comparing (multi) strange baryons and antibaryons of same type and also comparing different types with each other, e.g., $\Lambda,\ \Xi$ and $\Omega$~\cite{Abatzis:1991ju,Antinori:2001qk}. There is a steady rise of $s/S$ with energy  and centrality --- but perhaps for preliminary LHC data, and there is the predicted  enhancement of multi-strange hadrons and strange antibaryons as noted above.

Is the particle source that does all this indeed a QGP drop? All of the above  requires strange quark mobility. The chemical characteristics (non-equilibrium of hadron yield)  are consistent with sudden hadron production in fast breakup of QGP. Similarly, the enhanced source of entropy content is consistent with initial state  thermal gluon degrees of freedom,  which, in turn, was expected given strangeness enhancement.

To conclude, we can use (strange) hadron yields to learn about QGP properties at hadronization 
--- remote `sensing'. Strangeness fingerprints properties of QGP  and demonstrates deconfinement.

\section*{Acknowledgments:}
I am deeply indebted to Rolf Hagedorn whose continued mentoring 30+ years ago provided much of the guidance and motivation in my decades long pursuit of strangeness in quark--gluon plasma, a feat in which I imitated his constancy in dealing with disbelievers of elementary particle thermodynamics.  I thank my friends, colleagues, students: (alphabetically)  Marek Ga\'zdzicki, Peter Koch, Jean Letessier, Berndt Muller,  Emanuele Quercigh, Giorgio Torrieri who in past 30 years  have  been instrumental in developing and shaping the  understanding of strangeness signature of QGP. Among my academic `teachers', I am particularly indebted to John S. Bell, Iwo Bialynicki-Birula, Peter Carruthers, John W. Clark, Michael Danos, Walter Greiner, Maurice Jacob and Abraham Klein. I would like to thank friends and colleagues  who in  a friendly and constructive manner have been contributing important insights to the advancement of the subject, here, in particular, I wish to mention Federico Antinori, Tamas Biro, Marcus Bleicher, Wojtek Broniowski, Laszlo Csernai, Carsten Greiner, Hans Gutbrod, Huan Z. Huang, Joe Kapusta, Krzysztof Redlich, Dirk Rischke, Karel Safarik, Horst St\"ocker, Ludwik Turko.  I am deeply appreciative, and  indebted, to my friends Andrzej Bialas and Wojtek Florkowski for many years of friendship and fruitful collaboration, and for the organization of this magnificent Strangeness in Quark Matter (SQM2011) meeting in Krak\'ow, and the jubilee that came along with it. 

This work has been supported by a grant from the U.S. Department of Energy, DE-FG02-04ER41318.
\vskip -0.2cm

%

\end{document}